\documentstyle[12pt]{article}

\input epsf
\begin{document}
\baselineskip=20.5pt

\def\beqra{\begin{eqnarray}} \def\eeqra{\end{eqnarray}}
\def\beqast{\begin{eqnarray*}}
\def\eeqast{\end{eqnarray*}}
\def\beq{\begin{equation}}      \def\eeq{\end{equation}}
\def\be{\begin{enumerate}}   \def\ee{\end{enumerate}}

\def\fnote#1#2{\begingroup\def\thefootnote{#1}\footnote{
#2}
\addtocounter
{footnote}{-1}\endgroup}

\def\itp#1#2{\hfill{NSF-ITP-{#1}-{#2}}}

\def\gam{\gamma}
\def\Gam{\Gamma}
\def\la{\lambda}
\def\eps{\epsilon}
\def\La{\Lambda}
\def\si{\sigma}
\def\Si{\Sigma}
\def\al{\alpha}
\def\Tha{\Theta}
\def\tha{\theta}
\def\vphi{\varphi}
\def\del{\delta}
\def\Del{\Delta}
\def\ab{\alpha\beta}
\def\om{\omega}
\def\Om{\Omega}
\def\mn{\mu\nu}
\def\mun{^{\mu}{}_{\nu}}
\def\kap{\kappa}
\def\rsi{\rho\sigma}
\def\beal{\beta\alpha}

\def\til{\tilde}
\def\htil{\tilde{H}}
\def\rta{\rightarrow}
\def\eqv{\equiv}
\def\nab{\nabla}
\def\pa{\partial}
\def\sit{\tilde\sigma}
\def\ul{\underline}
\def\indt{\parindent2.5em}
\def\nd{\noindent}

\def\rsi{\rho\sigma}
\def\beal{\beta\alpha}

\def\caa{{\cal A}}
\def\cb{{\cal B}}
\def\cac{{\cal C}}
\def\cd{{\cal D}}
\def\ce{{\cal E}}
\def\cf{{\cal F}}
\def\cg{{\cal G}}
\def\cah{{\cal H}}
\def\ci{{\cal I}}
\def\cj{{\cal{J}}}
\def\ck{{\cal K}}
\def\cl{{\cal L}}
\def\cm{{\cal M}}
\def\cn{{\cal N}}
\def\cO{{\cal O}}
\def\cp{{\cal P}}
\def\car{{\cal R}}
\def\cs{{\cal S}}
\def\ct{{\cal{T}}}
\def\cu{{\ca{U}}}
\def\cv{{\cal{V}}}
\def\cw{{\cal{W}}}
\def\cx{{\cal{X}}}
\def\cy{{\cal{Y}}}
\def\cz{{\cal{Z}}}

\def\raisenot{\raise .5mm\hbox{/}}
\def\nota{\ \hbox{{$a$}\kern-.49em\hbox{/}}}
\def\notA{\hbox{{$A$}\kern-.54em\hbox{\raisenot}}}
\def\notb{\ \hbox{{$b$}\kern-.47em\hbox{/}}}
\def\notB{\ \hbox{{$B$}\kern-.60em\hbox{\raisenot}}}
\def\notc{\ \hbox{{$c$}\kern-.45em\hbox{/}}}
\def\notd{\ \hbox{{$d$}\kern-.53em\hbox{/}}}
\def\notbd{\ \hbox{{$D$}\kern-.61em\hbox{\raisenot}}} 
\def\note{\ \hbox{{$e$}\kern-.47em\hbox{/}}}
\def\notk{\ \hbox{{$k$}\kern-.51em\hbox{/}}}
\def\notp{\ \hbox{{$p$}\kern-.43em\hbox{/}}}
\def\notq{\ \hbox{{$q$}\kern-.47em\hbox{/}}}
\def\notW{\ \hbox{{$W$}\kern-.75em\hbox{\raisenot}}}
\def\notz{\ \hbox{{$Z$}\kern-.61em\hbox{\raisenot}}}
\def\notpa{\hbox{{$\partial$}\kern-.54em\hbox{\raisenot}}}

\def\fo{\hbox{{1}\kern-.25em\hbox{l}}}  
\def\rf#1{$^{#1}$}
\def\bx{\Box}
\def\tr{{\rm Tr}}
\def\rmtr{{\rm tr}}
\def\dgg{\dagger}

\def\lag{\langle}
\def\rag{\rangle}
\def\bmid{\big|}
\def\pw{P\left(w\right)}

\def\vlap{\overrightarrow{\La p}} 
\def\lrta{\longrightarrow}
\def\lrar{\raisebox{.8ex}{$\longrightarrow$}}
\def\rlarw{\longleftarrow\!\!\!\!\!\!\!\!\!\!\!\lrar}

\def\llra{\relbar\joinrel\longrightarrow}     
\def\mapright#1{\smash{\mathop{\llra}\limits_{#1}}}
\def\mapup#1{\smash{\mathop{\llra}\limits^{#1}}}
\def\asymptotic{{_{\stackrel{\displaystyle\longrightarrow}
{x\rightarrow\pm\infty}}\,\, }} 
\def\asymptext{\raisebox{.6ex}{${_{\stackrel{\displaystyle\longrightarrow}
{x\rightarrow\pm\infty}}\,\, }$}} 

\def\7#1#2{\mathop{\null#2}\limits^{#1}}   
\def\5#1#2{\mathop{\null#2}\limits_{#1}}   
\def\too#1{\stackrel{#1}{\to}}
\def\tooo#1{\stackrel{#1}{\longleftarrow}}
\def\nout{{\rm in \atop out}}

\def\one{\raisebox{.5ex}{1}}
\def\BM#1{\mbox{\boldmath{$#1$}}}

\def\ltsim{\matrix{<\cr\noalign{\vskip-7pt}\sim\cr}}
\def\gtsim{\matrix{>\cr\noalign{\vskip-7pt}\sim\cr}}
\def\haf{\frac{1}{2}}


\def\place#1#2#3{\vbox to0pt{\kern-\parskip\kern-7pt
                             \kern-#2truein\hbox{\kern#1truein #3}
                             \vss}\nointerlineskip}

\def\illustration #1 by #2 (#3){\vbox to #2{\hrule width #1
height 0pt
depth
0pt
                                       \vfill\special{illustration #3}}}

\def\scaledillustration #1 by #2 (#3 scaled #4){{\dimen0=#1
\dimen1=#2
           \divide\dimen0 by 1000 \multiply\dimen0 by #4
            \divide\dimen1 by 1000 \multiply\dimen1 by #4
            \illustration \dimen0 by \dimen1 (#3 scaled #4)}}

\def\ON{{\cal O}(N)}
\def\UN{{\cal U}(N)}
\def\bdPh{\mbox{\boldmath{$\dot{\!\Phi}$}}}
\def\bPh{\mbox{\boldmath{$\Phi$}}}
\def\bPhs{\bPh^2}
\def\sef{S_{eff}[\sigma,\pi]}
\def\sigx{\sigma(x)}
\def\pix{\pi(x)}
\def\bph{\mbox{\boldmath{$\phi$}}}
\def\bphs{\bph^2}
\def\ex{\BM{x}}
\def\exs{\ex^2}
\def\xdot{\dot{\!\ex}}
\def\y{\BM{y}}
\def\ys{\y^2}
\def\ydot{\dot{\!\y}}
\def\pat{\pa_t}
\def\pax{\pa_x}

\renewcommand{\theequation}{\arabic{equation}}


\itp{97}{142}\\

\vspace*{.3in}
\begin{center}
 \large{\bf A Non-Hermitean Particle in a Disordered World*}

\vspace{36pt}
{A. Zee}
\end{center}

\vskip 1mm
\begin{center}
{Institute for Theoretical Physics,}\\
{University of California, Santa Barbara, CA 93106, USA}
\vspace{.6cm}

\end{center}

\begin{minipage}{5.3in}
{\abstract~~~~~ There has been much recent work on the spectrum of the
random non-hermitean Hamiltonian which models the physics of vortex line
pinning in superconductors. This note is loosely based on the talk I gave
at the conference ``New Directions in Statistical Physics" held in Taipei,
August 1997. We describe here new results in spatial dimensions higher than
one. We also give an expression for the spectrum within the WKB approximation.}
\end{minipage}

\vspace{48pt}
\vfill
\vspace{60pt}
{\rule{2in}{.01in}\\
{\small *To appear in the proceedings of ``New Directions in Statistical Physics," Taipei, August 1997, Physica A, to be published.}
\pagebreak

\setcounter{page}{1}

\section{Introduction}
A non-hermitean Hamiltonian inspired by the problem of vortex line pinning
in superconductors has attracted the attention of a number of authors
\cite{hatano, hn2, efetov, feinzee, zahed, beenakker, zee, bz, gk, spectral}. In one
version, we are to study the eigenvalue problem

\beq\label{eig}
\sum_j~H_{ij} \psi_j ~=~ {t \over 2} \left(e^h ~ \psi_{i+1} ~+~
e^{-h}~\psi_{i-1}\right) ~+~ w_i \psi_i ~=~ E\psi_i, ~i,j ~=~ 1, \ldots, N
\eeq
with the periodic identification $i+N\equiv i$ of site indices. Here the
real numbers $w_i$'s, the site energies, are drawn independently from some probability
distribution $P\left(w\right)$ (which we will henceforth take to be even
for simplicity.)
This Hamiltonian describes a particle hopping on a ring, with its clockwise
hopping amplitude different from its counter-clockwise hopping amplitude.
On each site there is a random potential. The number of sites $N$ is
understood to be tending to infinity. We will also take $t$ and $h$ to be
positive for definiteness. (The hopping amplitude $t$ can be scaled to $2$
for instance but we will keep it for later convenience.) 

Note that the
Hamiltonian is non-hermitean for $h$ non-zero and thus has complex
eigenvalues. It
is represented by a real non-symmetric matrix, with the reality implying
that if $E$ is an eigenvalue, then $E^*$ is also an eigenvalue. For $\pw$
even, for each particular realization $\{w_i\}$ of the random site
energies, the realization $\{-w_i\}$ is equally likely to occur, and thus
the spectrum of $H$ averaged with $\pw$ has the additional symmetry $E ~\rta~ -E$.

Without non-hermiticity ($h=0$), all eigenvalues are of course real, and
Anderson and collaborators \cite{anderson} showed that all states are
localized. Without impurities ($w_i=0$), Bloch told us that
the Hamiltonian is immediately solvable with the eigenvalues
\beq\label{spectrum}
E(\theta) ~=~ t~{\rm cos}~( \theta - ih)~=~ {\rm cos}~\theta ~{\rm cosh} ~h ~+~ i ~{\rm sin} ~\theta ~{\rm sinh}~h \,, 
\eeq
with $\theta={2\pi n\over N}, \quad (n = 0, 1, \cdots, N-1)$,
tracing out an ellipse. The corresponding wave functions $\psi^{(n)}_j\sim~e^{i \theta j}$ are obviously extended. We are to study what
happens in the presence of both non-hermiticity and the impurities.

\begin{figure}[htbp]
\epsfxsize=4in
\begin{center}
\leavevmode
\epsfbox{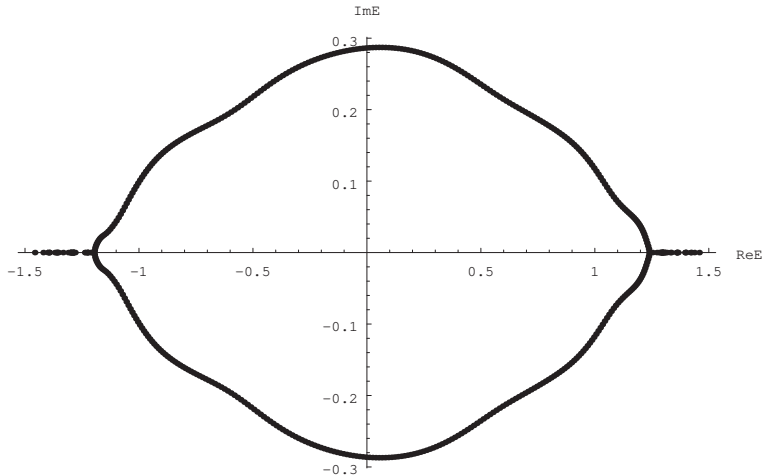}
\end{center}
\caption{The eigenvalues of one particular realization of the Hamiltonian in (1) with $N=500$, $t=1$, $h=0.4$ and with the random site energies $w_i$ equal to $\pm 0.5$.}
\label{fig1}
\end{figure}

In figure (1) we show a typical spectrum for the Hamiltonian in (\ref{eig}).
The spectrum is made of two parts, a more or less elliptical (or ``lemon-like")
curve, consisting of complex eigenvalues, and ``two wings," a right wing and a
left wing, consisting of real eigenvalues, poking out of the complex part of
the spectrum.

In the literature \cite{hatano, hn2, ld, dg, mf} it was argued that localized states have real eigenvalues 
and that states of $H$ corresponding to complex eigenvalues are extended,
that is, delocalized. Remarkably, non-hermitean localization theory is
simpler in this respect than the standard hermitean localization theory of
Anderson and others \cite{anderson}. To understand the localization
transition, we have to study only the energy spectrum of $H$.

I refer the reader to the talk by N. Hatano in these proceedings for an
illuminating discussion of this subject.

In the talk at the conference, I described work done with J. Feinberg \cite{zee, spectral} and
with E. Br\'ezin \cite{bz}. Since the material covered can be found
discussed in detail in these references, it may be more useful to the
reader for me to describe here some slight and unpublished work done with
E. Br\'ezin. We will mention two separate topics here.

\pagebreak

\section{Continuum Limit and $WKB$ Approximation}

As is obvious and as has been discussed in the literature, we can easily
take the continuum limit of (\ref{eig}) by letting $x~=~ia$, $i ~\rta~
\infty$, $a~\rta~0$, $t ~\rta~ \infty$, $ta^2 ~=~ -{1 \over m}$, $\eps ~=~
E~-~t$, and $A ~=~ {h \over a}$. Note that we have introduced the gauge
potential $A$, equal to $h$ divided by the lattice spacing, and hence with
the usual dimension of inverse length. We obtain

\beq\label{con}
-{1 \over 2m} ~\left[(\nabla ~+~ A)^2 ~-~ A^2 \right] ~\psi ~+~ V \psi ~=~
\eps~\psi
\eeq
Inserting $\psi ~=~ e^{iS}$ we have (setting $2m ~=~ 1$)

\beq\label{ess}
S'^2 ~-~ 2iAS' ~-~ iS'' ~=~ \eps ~-~ V
\eeq
The $WKB$ approximation involves neglecting $S''$, as usual. We obtain

\beq\label{ess2}
S' ~=~ iA ~\pm~ \sqrt{-A^2 ~+~ (\eps ~-~ V)}
\eeq
with the boundary condition

\beq\label{ess3}
S(L) ~=~ S(0) ~+~ 2 \pi ~n
\eeq
Let us define

\beq\label{kay}
k ~\equiv~ {2\pi~n \over L}
\eeq
In the limit $L~\rta~\infty$, $k$ becomes a continuous variable. Then we find

\beq\label{kay2}
(k ~-~ iA) ~=~ \pm~ {1 \over L} ~ \int_0^L ~ dx ~ \sqrt{\eps ~-~ A^2 ~-~ V}
\eeq

One way to look at this expression is to study the effects of the
non-hermitean potential $A$. In the absence of the non-hermiticity, we have

\beq\label{herm}
k ~=~ \pm~ {1 \over L} ~ \int_0^L ~ dx ~ \sqrt{\eps_0 (k) ~-~ V}
\eeq
We thus obtain a remarkably compact expression for the $WKB$ spectrum of
(\ref{con})

\beq\label{compact}
\eps(k) ~=~ \eps_0 ~(k ~-~ iA) ~+~ A^2
\eeq
Note that we have to know the usual hermitean spectrum $\eps_0(q)$ for
complex values of its argument. In particular, we can study the effect of
non-hermiticity perturbatively

\beq\label{pert}
\epsilon(k) ~=~ \eps_0(k) ~-~ iA ~ {d \eps_0 \over dk} ~+~ A^2
\eeq
The expressions (\ref{compact}) and (\ref{pert}) show clearly how 
non-hermiticity drives the eigenvalues into the complex plane. For instance, if 
$\epsilon_0 (k)~=~\epsilon_0~+~ak^2~+~bk^4~+~\ldots$, then $\epsilon 
(k)~=~\epsilon_0~+~A^2(1-a) ~-~2 i a ~Ak ~+~ak^2 ~+~ \ldots$.

This discussion holds whether $V$ is random or not. For random $V$, it
would be interesting to see if it would be possible to obtain a relation
between the non-hermitean density of states

\beq\label{denstoo}
\rho(x,y) ~=~ \lag~\int ~dk ~ \delta \left(x~-~A^2 ~-~ Re ~\eps_0 (k~-~iA)
\right) ~\delta \left( y ~-~ Im ~ \eps_0 ~(k~-~iA) \right) \rag
\eeq
and the hermitean density of states

\beq\label{denstoo2}
\rho(\mu) ~=~ \lag ~\int ~dk ~\delta \left(\mu ~-~ \eps_0(k) \right) \rag
\eeq
\pagebreak

\section{Higher Dimensions}

Much of the literature on this subject is numerical and, with some
exceptions, focused on one dimensional space. In real systems,
vortices move in two dimensional space. In a recent work \cite{ns}, Nelson
and Shnerb studied the higher dimensional problem using numerical,
perturbative, renormalization group, and other techniques. Thus, it may be
of some interest to report here some exact analytic results in higher
dimensions.

First, we have to review some basic formalism. As is standard, we are to
study the Green's function

\beq\label{green}
G(z) = {1 \over z-H}
\eeq
The averaged density of eigenvalues, defined by

\beq\label{master}
\rho(x,y) ~\equiv~ \langle {1\over N} \sum_i \delta(x-{\rm Re} E_i)~ \delta
(y- {\rm Im}
E_i)\rangle \,,
\eeq
is then obtained (upon recalling the identity ${\partial \over \partial
z^\ast} ~ {1 \over z} ~=~ \pi \delta(x) ~\delta(y)$) as

\beq\label{master2}
\rho\left(x,y\right) ~=~ {1 \over \pi} {\partial \over \partial z^\ast}
\overline{G\left(z\right)}
\eeq
with $z=x+iy$. Here we define

\beq\label{master3}
\overline{G\left(z\right)} \equiv \lag \frac{1}{N}~ {\rm tr}~ G
\left(z\right) \rag
\eeq
and $< \ldots >$ denotes, as usual, averaging with respect to the
probability ensemble from which $H$ is drawn. It is also useful to define
the ``bare" Green's function

\beq\label{gdis}
G_0(z) = {1 \over z-H_0}
\eeq
in the absence of impurities and the corresponding $\overline{G_0(z)}
\equiv {1 \over N} ~{\rm tr} ~ G_0(z)$. For the spectrum in
(\ref{spectrum}), we can determine $\overline{G_0(z)}$ explicitly \cite{bz}. We need hardly remark that in determining $\overline{G_0(z)}$ no averaging is involved.

In \cite{bz} a formula relating the Green's function $G$ to $G_0$ was
obtained. Here we give an alternative derivation, perhaps somewhat more
direct, of this relation. From
\beq\label{ham}
H ~=~ H_0 ~+~ \sum\limits_k ~w_k ~ P_k
\eeq
where we defined the
projection operator
\beq
P_k ~=~ \mid k \rag \lag k \mid
\eeq
onto the site $k$, and
the definitions of $G$ and $G_0$ we have

\beq\label{greentoo1}
G^{-1} ~=~ G^{-1}_0 ~-~ \sum\limits_k ~w_k ~P_k ~=~ G^{-1}_0 (1~-~
\sum\limits_k~ G_0 ~w_k ~P_k)
\eeq
and hence

\beq\label{greentoo2}
G ~=~ \left( {1 \over 1~-~ \sum\limits_k~ G_0 ~w_k ~P_k} \right) ~G_0
\eeq
We might consider expanding this expression as a power series in the
$w_k$'s and then average with $P(w)$. Indeed, that would be the correct
procedure for many distributions. However, for some $P(w)$ such
as the Cauchy distribution
\beq\label{Cauchy}
P(w) = {\gam\over \pi} {1\over w^2+\gam^2}\,,
\eeq
with its long tails extending to infinity, the moments $\lag w_k^n \rag$,
for $n$ an integer, do not exist. The physics here is that we have to first sum the
effect of repeated scattering on the impurity
potential at one particular site $k$

\beq\label{summation}
\upsilon_k ~\equiv~ {w_k \over 1-w_k \lag k \mid G_0 \mid k \rag} ~=~ {w_k
\over 1 ~-~w_k ~ \overline{G}_0}
\eeq
The last equality follows from the assumed translation invariance of
$H_0$. Since the probability distribution $P(w)$ is of course normalizable,
the moments $\lag \upsilon _k^n \rag$,
for $n$ an integer, in contrast to $\lag w_k^n \rag$,
always exist. Note that in
contrast to $w_k$, $\upsilon_k$ is a function of $z$. We can solve for
$w_k$ in terms of $\upsilon_k$ as

\beq\label{wkay}
w_k ~=~ {\upsilon_k \over 1 ~+~ \upsilon_k ~ \overline{G_0}}
\eeq
Inserting (\ref{wkay}) into (\ref{greentoo1}) we obtain
without further ado

\beq\label{together}
G~=~ \left[1-\sum_k\left(\frac{\upsilon_k G_0 P_k}{1+\upsilon_k G_0 P_k}
\right) \right]^{-1} ~ G_0 ~=~\left[1-\sum_k \frac{\upsilon_k G_0
P_k}{1+\upsilon_k \overline{G_0}} \right]^{-1} ~ G_0
\eeq
The reader is invited to average the trace of (\ref{together}) with
his or her favorite $\pw$, and thence to obtain the density of eigenvalues
$\rho(x,y)$ by (\ref{master2}).

In general, it is non-trivial to carry out this averaging explicitly. For
the Cauchy distribution, however, we have the drastic simplification that

\beq\label{moments}
\lag \upsilon_k^n \rag ~=~ \lag \upsilon_k \rag^n
\eeq
with

\beq\label{frist}
\lag \upsilon_k \rag ~=~ 
{i~\gam \over 1~-~i \gam ~ \overline{G}_0} ~\,, {\rm for~ Im}~ z ~<~ 0 
\eeq
and

\beq\label{second}
\lag \upsilon_k \rag ~=~ {-i~\gam \over 1~+~i \gam ~ \overline{G_0}} ~\,, ~{\rm for~ Im}~ z ~>~ 0
\eeq
which by translation invariance is independent of $k$. Defining as usual

\beq\label{sig}
\lag G \rag^{-1} ~=~ G_0^{-1} ~-~ \Sigma(z)
\eeq
we see that for the Cauchy distribution

\beq\label{sig2}
\Sigma(z) ~=~ {\lag \upsilon \rag \over 1~+~ \lag \upsilon \rag ~
\overline{G}_0}
\eeq
and is thus equal to $\mp i \gamma$ in the
upper and lower half plane respectively. We thus obtain from (\ref{master3})

\beq\label{res}
\overline{G(z)} ~=~ \overline{G_0(z-\Sigma)}
\eeq

Contrary to some erroneous claim in the literature, for an arbitrary probability
distribution $\Sigma(z)$ is in
general a matrix and (\ref{res}) does not hold. 

For the Cauchy distribution, however, we obtain from (\ref{sig2}) and (\ref{res})

\beq\label{L}
\overline{G(z)} ~=~ \overline{G_0 \left( z + i \gamma \right)} ~ \theta
\left( {\rm Im} ~z \right) ~+~ \overline{G_0 \left(z-i \gamma \right)} ~
\theta\left({\rm -Im}~ z \right)
\eeq

The density of  eigenvalues are given by (\ref{master2}).
When
${\partial \over \partial z^{\ast}}$ in (\ref{master2}) acts on
$\overline{G_0}$ in (\ref{L}) we obtain the ``central part" of the spectrum

\beq\label{M}
\rho_{central}(x,y) ~=~ \rho_0(x,y + \gamma) ~ \theta(y) ~+~ \rho_0(x, y- \gamma) ~
\theta(-y)
\eeq
In other words, the random site energies push the density of eigenvalues
$\rho_0(x,y)$ of the non-random problem towards the real
axis by a distance $\gamma$. This accounts for the complex eigenvalues. In
figure (1) this central part of the spectrum consists of a curve \cite{gk, spectral}. However, in general, in higher dimensional space for instance, the central part
of the spectrum forms a two dimensional blob.

When ${\partial \over \partial z^{\ast}}$ in
(\ref{master2}) acts on the step functions in (\ref{L}) we obtain the
density on the two ``wings":

\beq\label{Ngen}
\rho_{wing}(x,y) ~=~ -{1 \over 2 \pi i} ~ \delta (y) ~\left(\overline{G_0 (x+i \gam)}~-~ \overline{G_0 (x-i \gam)}\right)~=~ \delta (y) {\gam \over \pi} ~{1 \over N}~
tr ~{1 \over (x - H_0)^2 ~+~ \gam^2}
\eeq
We thus have the general result that $\rho_{wing}$ is constructed out of the sum of Cauchy distributions centered about each of the eigenvalues of $H_0$. At first sight, this result seems puzzling, since $\rho_{wing}$ is not
obviously real. The resolution is that we have to look more closely at how
the averaging over the Cauchy distribution is done. The discussion in the
literature \cite{ziman} is given for hermitean $H_0$'s, and hence
had to be generalized to non-hermitean $H_0$'s in \cite{bz}.  Imagine
expanding (\ref{together}) in a series in the $\upsilon_k$'s. According to
(\ref{moments}) we have to average $\upsilon={w\over w-{\overline G_0(z)}}$, that is, to
integrate $\upsilon P(w)$ over $w$. In \cite{bz}, $\overline {G_0(z)}$ is explicitly known, and
we can see that for $z$ in the upper half plane, the integration contours
in $w$ can be closed in the lower half plane and by Cauchy's theorem $w$ is
effectively set
equal to $-i \gamma$, and vice versa if $z$ is in the lower half plane.

In the present discussion, we have to find a class of $H_0$'s for which
this procedure continues to hold. It is not too difficult to find such a
class. Suppose the eigenvalue spectrum of $H_0$ is such that if $E$ is an
eigenvalue, then $E^*$ is also an eigenvalue. As was mentioned in the
introduction, this is the case if $H_0$ is real (but not necessarily
symmetric.) Then $\overline{G_0(z)}^* ~=~ \overline{G_0(z^*)}$ and it is
straightforward to show that the integration procedure
outlined above holds. It is an interesting question whether there exists a
class of $H_0$'s broader than the one found here for which this integration procedure still holds.

We now see, gratifyingly enough, that for the class of $H_0$'s found here
$\rho_{wing}$ is indeed real and positive definite.  Indeed, we can write
(\ref{Ngen})
entirely in terms of $\rho_0(x,y)$:

\beq\label{tom}
\rho_{wing}(x,y) ~=~ \delta (y) ~ {\gamma \over \pi} \int dudv \rho_0(u,v)
{(x-u)^2 - v^2 + \gamma^2 \over [(x-u)^2 - v^2 + \gamma^2]^2 +
4v^2(x-u)^2}
\eeq
where we used $\rho_0(u,v)=\rho_0(u,-v)$.

Thus, we simply have to evaluate the Green's function for a given $H_0$, plug
in (\ref{M}) and (\ref{Ngen}), and obtain the density of states for $H$. We
emphasize that in this derivation the only thing we assumed for $H_0$ is its
translation invariance. Our result applies even if $H_0$ contains non-nearest
neighbor hopping, for instance. In particular, this result holds for any spatial
dimension. 

As an example, in \cite{bz} we applied this result to the $H_0$ in (\ref{eig})
and obtained 
\beq\label{N} \rho_{wing}(x,y) ~=~ {1 \over \sqrt{2} \pi} ~ \delta
(y) ~ \theta (x^2-x^2_{min} (\gamma)) ~ {\sqrt{\gamma^2-x^2+t^2 ~+~
B(x,\gamma,t)} \over B(x, \gamma, t)} 
\eeq 
where we have defined 
\beq 
B(x,\gamma, t) ~\equiv~ \sqrt{(x^2 + \gamma^2)^2 ~+~ 2t^2(\gamma^2-x^2) ~+~ t^4}
\eeq 
and

\beq\label{conv}
x_{min}(\gamma) ~\equiv~ {(\sqrt{(t ~{\rm sinh}~ h)^2 ~-~ \gamma^2}) \over {\rm
tanh}~ h}
\eeq
The critical value of $\gamma$ at which the wings extend over the entire real axis is determined by $x_{min}(\gamma)=0$:

\beq\label{critical}
\gamma_c ~=~ t ~{\rm sinh}~ h
\eeq
For randomness larger than $\gamma_c$, all states are localized.

Note that the density of eigenvalues on the wings does not depend on the value of the non-hermiticity $h$ at all; $h$ comes in only in determining where the wings end. This explicitly verifies the argument \cite{hatano, hn2, ld, dg, mf} in the literature that 
the localized states
do not know about the non-hermiticity.

We now understand that most of the qualitative features exhibited by these
results obtained in \cite{bz} are generic for the Cauchy distribution.  From (\ref{Ngen}), we
see that $\gam_{c1}$, defined as the critical randomness at which wings first
appear, is equal to $0^+$ for any translation invariant $H_0$.  We also obtain a
general result for $\gamma_c$, the critical randomness at which all states are
localized, namely
\beq\label{ymx}
\gamma_c = y_{max}
\eeq
where $y_{max}$ is the maximum value of $y$ in the support of $\rho_0(x,y)$.

In summary, in any dimensions, for any $H_0$ whose eigenvalues come in complex
conjugate pairs, and with the random site energies governed by the Cauchy
distribution, we have the result that the density of eigenvalues is given by
the sum of two terms, $\rho_{central}(x,y)$ in (\ref{M}), and
$\rho_{wing}(x,y)$ in (\ref{Ngen}) which has the form $\delta(y) ~f(x)$. By the way,
this explicit delta funtion accounts for the ``ridge" along the real axis seen
in the numerical work of Hatano and Nelson (see figure (19c) in \cite{hn2}). In the numerical result, we see a background consisting of a two dimensional
blob on which eigenvalues are more or less uniformly distributed. 
Superimposed on this background is a dense cluster of eigenvalues on the real
axis, the so called ``ridge."  (The entire spectrum is thus reminiscent of a
cross section of a galaxy, with a halo and a concentration of stars in the
galactic plane.)

The numerical work of Hatano and Nelson was actually performed for the box
distribution $P(w) = {1\over {2V}} \theta(V^2-w^2)$.  We thus have an indication
that some features of the spectrum are generic.  One feature which can clearly
be attributed to the fact that the Cauchy distribution has a long tail is that
our $\rho_{wing}(x,y)$ in (\ref{Ngen}) extends to $x = \pm\infty$, falling off as $1\over{x^2}$.  Invoking the galactic analogy, we can describe the situation
for Cauchy distribution by saying that the halo is concentrated near the center
of the galactic plane, with the density of stars thinning out rapidly, while
for distribution with finite support the halo more or less fits over the
galactic plane.

A quantity of some interest is the fraction of eigenvalues in the ``galactic
plane", which we can obtain by integrating $\rho_{wing}(x,y)$.  We find
\beq\label{wing}
\int^{\infty}_{-\infty}\rho_{wing}(x,y) = \delta(y) \int dudv \rho_0(u,v)
\theta(-\gamma<v<\gamma)
\eeq
Note that $\int\limits_{- \infty}^\infty ~dx ~\rho_{wing}~(x,y) = \delta(y)$ for $\gam$ larger than $\gam_c$, in
accordance with (\ref{ymx}).

The reader is invited to evaluate $G_0(z)$ for his or her favorite $H_0$.

We will give two illustrative examples here. 
As a first simple example, we can consider a ladder, on whose two legs are two 
copies of the Hamiltonians in (\ref{eig}), and with hopping between
the two legs given by the amplitude $t'$. More precisely, we take

\beqra\label{lad}
H_{0 ~\alpha i, ~\beta j} ~= &{t \over 2} \left(e^h ~\delta_{\alpha
\beta}~\delta_{i+1,~j} ~+~ e^{-h} ~ \delta_{\alpha \beta}~ \delta_{i,j+1}
\right) \nonumber \\
&+ ~{t' \over 2}~\delta_{ij}~\left(\delta_{\alpha+1,~\beta} ~+~
\delta_{\alpha, ~\beta +1} \right)
\eeqra
where $\alpha,~\beta$ take on two values and denote the two legs. The
periodic boundary condition $i~=~i+N$ is imposed as before. Alternatively,
we can think of a particle carrying a binary label (spin, flavor, etc)
hopping on a ring. This model was studied numerically by Hatano. See these
proceedings.

The eigenvalues of $H_0$ are obtained immediately

\beq\label{elad}
E(\theta) ~=~ t~{\rm cos}~\left(\theta ~-~ ih \right) ~\pm~{t' \over 2}
\eeq
tracing out two overlapping ellipses separated by $t'$. The corresponding
$G_0$ is the sum of two pieces

\beq\label{sum}
G_0(z) ~=~ g_0 ~\left(z ~+~ {t' \over 2} \right) ~+ ~g_0 ~\left(z ~-~ {t'
\over 2} \right)
\eeq
where $g_0(z)$ is the Green's function for the spectrum in (\ref{spectrum})

Now we turn on Cauchy randomness for the site energies. According to (\ref{M}) and (\ref{Ngen}),
the spectrum is just the superposition two ``squashed ellipses," each with
its own wings poking out. For comparison, we show the numerical result in figure
(2).
\begin{figure}[htbp]
\epsfxsize=4in
\begin{center}
\leavevmode
\epsfbox{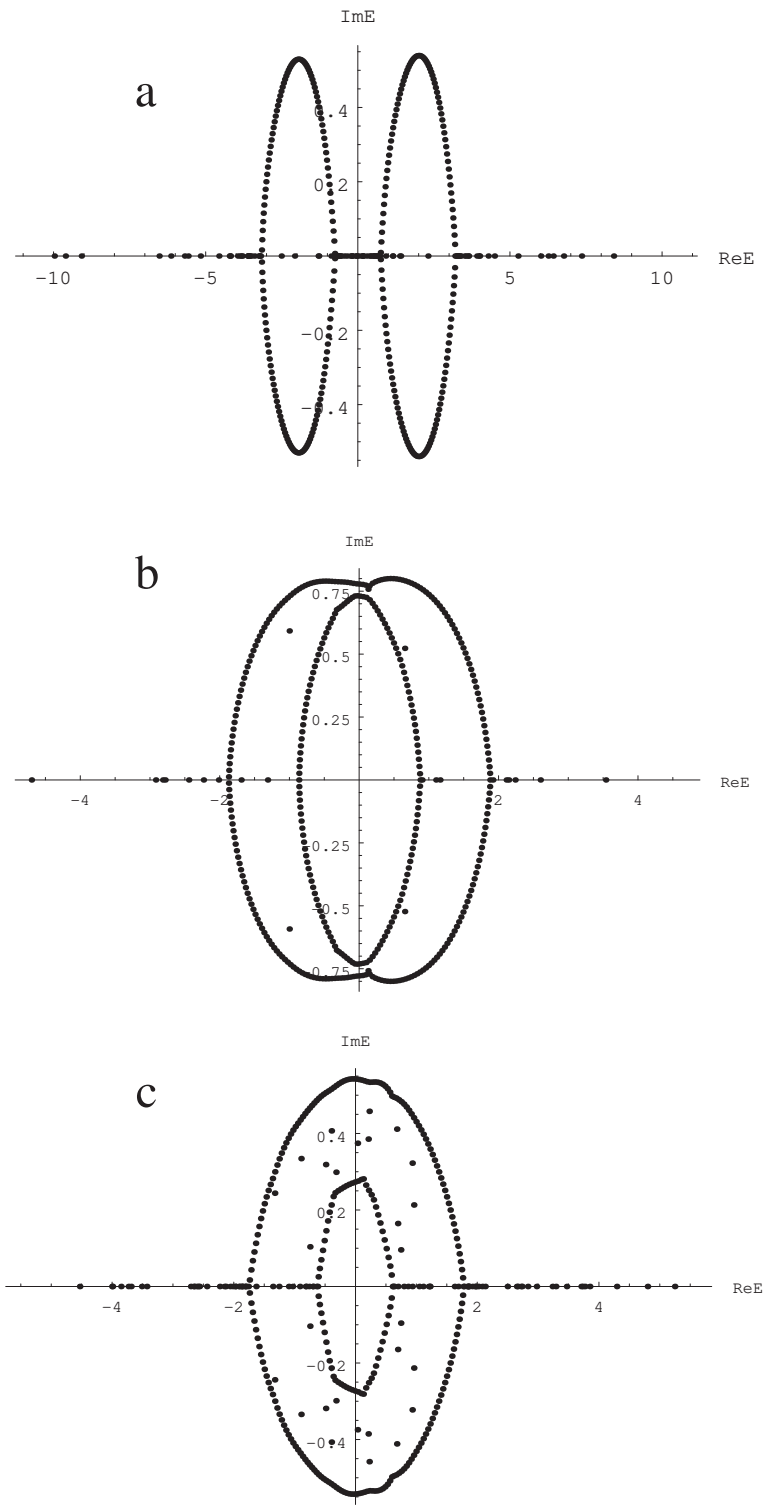}
\end{center}
\caption{The spectrum for the ladder problem (with $N~=~200$): (a) For $t'$ large (here $t' ~=~ 2$ with $t ~=~ 1$ and $h ~=~ 0.8$), the two ellipses are separated. The spectrum is shown for $\gam ~=~ 0.3$. (b) For smaller $t'$ (here $t' ~=~ 1$, with $t~=~
1$, $h~=~0.8$), the two ellipses overlap. The spectrum is shown for $\gam ~=~ 0.1$ (c) Same as (b) but with $\gam ~=~ 0.4$}
\label{fig2}
\end{figure}

We will finally venture into two dimensional space. We can of course immediately write down the analog of (\ref{eig}) for a lattice of any dimension. For two dimensions, we denote the parameters for hopping in the $x$ direction by ${t, h}$ and for hopping
 in the $y$ direction by ${t',h'}$.
The spectrum of $H_0$ is obviously
given by
\beq\label{spectrum2}
E(\theta,\theta')~ = ~t~{\rm cos}~( \theta - ih)+ t'~{\rm cos}~( \theta' - ih')\,, 
\eeq
with $\theta={2\pi n\over N},~ \theta'={2\pi n'\over N}, \quad (n,n' = 0, 1, \cdots, N-1)$. The spectrum is a two dimensional blob, constructed geometrically by placing an ellipse centered on each and every point of another ellipse. See figure (3). 

Depending on the values of $t$, $t'$, $h$, and $h'$, the spectrum can have different
topologies.  Let us refer to the case in which $t~=~t'$ as the isotopic hopping case.
Note that in that case the ``hole" in the spectrum closes when the hermiticity field
$(h,h')$ points in the $(\pm1, \pm1)$ direction.  (It is straightforward to take the
continuum limit.  In general, the mass $m$ in (\ref{con}) becomes anisotropic; in the
isotropic hopping case, we obtain (\ref{con}) with the replacement $A ~\rta~ \vec{A} ~=~ {1
\over a}(h,h')$.)

The corresponding Green's function  
\beq\label{onex}
G_0(z) ~=~ \int ~{d \theta \over 2 \pi }~ {d \theta ' \over 2 \pi} ~{1
\over z- E(\theta,\theta')}
\eeq
can be obtained by quadrature. We would not bother to carry out this computation here. Instead, for illustration, we will compute this integral only in a simplifying limit.
\begin{figure}[htbp]
\epsfxsize=4in
\begin{center}
\leavevmode
\epsfbox{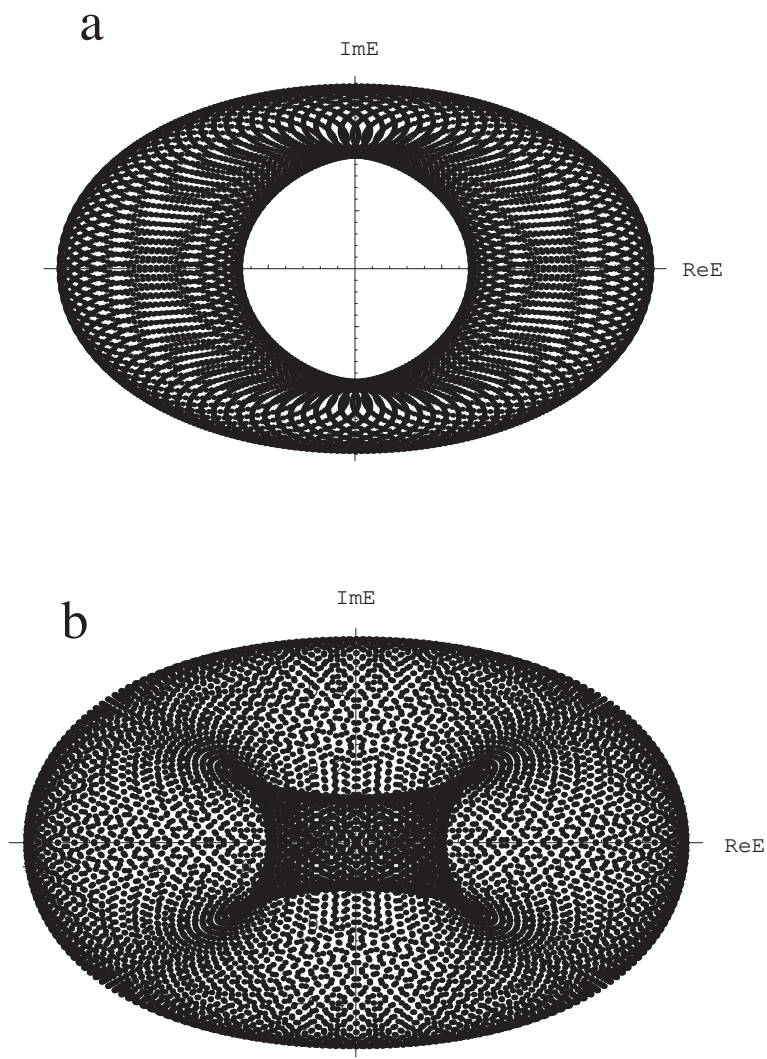}
\end{center}
\caption{The spectrum of $H_0$ in two dimensional space (ellipse + ellipse) for (a) $t=0.5$, $h=0.3$, $t' = 1$, $h'=0.6$ and for (b) $t=0.5$, $h=0.6$, $t' = 1$, $h'=0.2$.}
\label{fig2}
\end{figure}

In \cite{zee}, it was pointed out the the problem (\ref{eig}) simplifies in the
maximally non-hermitean or ``one way" limit, in which the parameters in
(\ref{eig}) are allowed to tend to the (maximally non-hermitean)
limit $h\rightarrow\infty$ and $t\rightarrow 0$ such that
\beq\label{thlim}
{t \over 2}~e^h\rightarrow 1
\eeq
The particle in (\ref{eig}) can only hop one way. The spectrum
(\ref{spectrum}) becomes
\beq\label{circle}
E(\theta) ~=~ e^{i \theta}\,, 
\eeq
with $\theta ~=~ {2 \pi n \over N} \,, (n = 0, 1, \cdots, N-1)$ and the ellipse associated with (\ref{spectrum}) expands into the unit
circle. The corresponding Green's function can be determined immediately
 
\beqra\label{oneone} G_0(z) &= &\int ~{d \theta \over 2 \pi } ~{1 \over z-e^{i
\theta}} \nonumber \\ &= &~{\theta \left(\mid z \mid^2 -1 \right) \over z}
\eeqra (We trust the reader not to confuse the Heaviside step function with the
angle variable.) For $\mid z \mid ~<~1$, $G_0(z) ~=~ 0$, and for $\mid z \mid
~>~1$, $G_0(z) ~=~ {1 \over z}$.  By (\ref{master2}) the  density of
eigenvalues vanishes in both regimes. The density of eigenvalues indeed has
support on the unit circle, in accordance with (\ref{circle}).
\begin{figure}[htbp] \epsfxsize=4in \begin{center} \leavevmode
\epsfbox{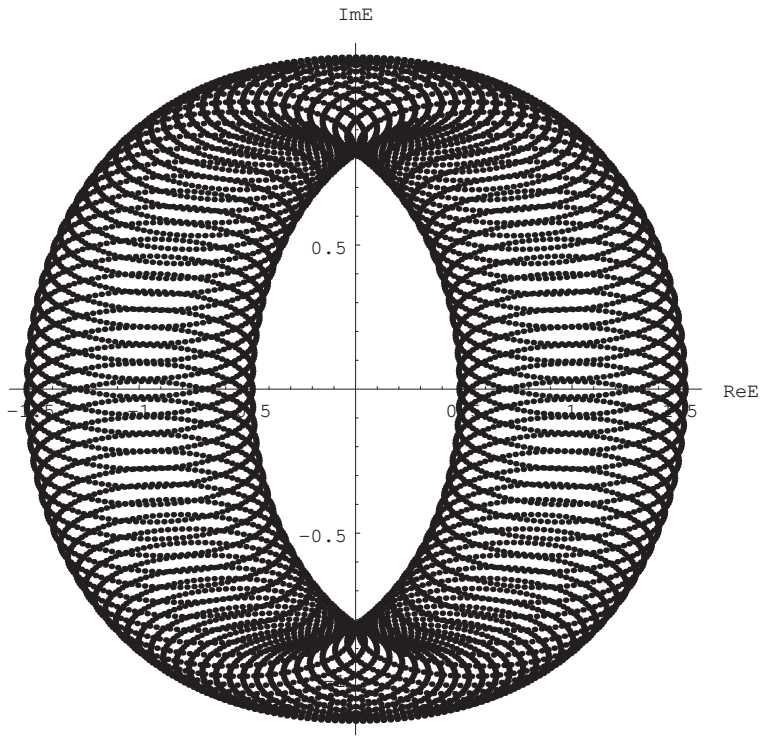} \end{center} \caption{The spectrum of $H_0$ in the single
``one way" limit for $t=0.5$, $h=0.3$ (ellipse plus circle).} \label{fig2}
\end{figure}

In two dimensional space, we can obviously take either the single ``one way" limit in which hopping along the $y$ direction becomes one way, or the double ``one way" limit in which hopping along both the $x$ and $y$ directions becomes one way. From (\ref{spectrum2}), we see that the spectrum changes from ``ellipse + ellipse" to ``ellipse $+$ circle" to ``circle $+$ circle." We show the spectrum for the single ``one way" limit in figure (4) and for the double ``one way" limit in figure(5).
\begin{figure}[htbp]
\epsfxsize=4in
\begin{center}
\leavevmode
\epsfbox{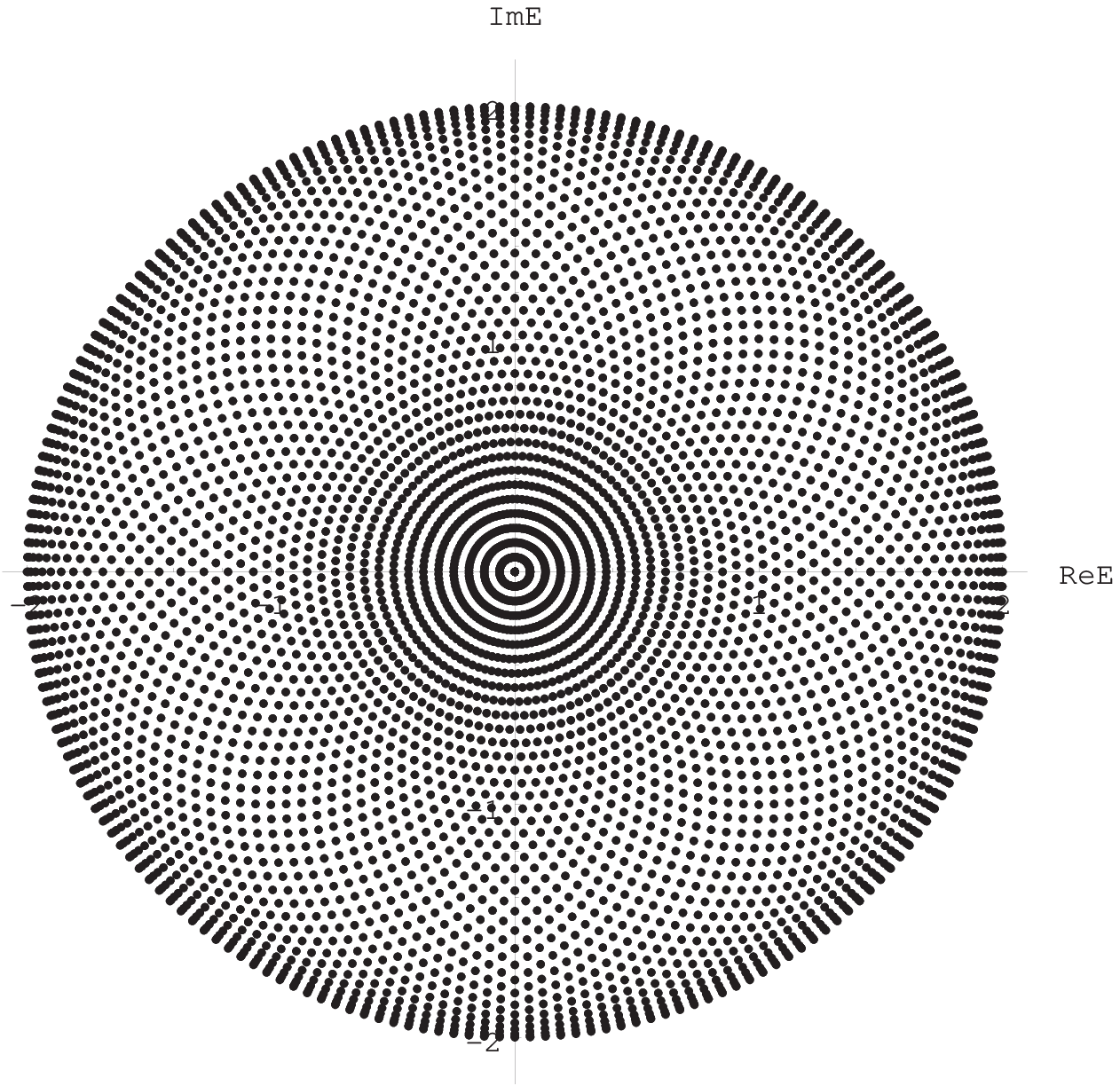}
\end{center}
\caption{The spectrum of $H_0$ in the double ``one way" limit (circle plus circle).}
\label{fig2}
\end{figure}

The double ``one way" limit is taken by letting 
$h, h'\rightarrow\infty$ and $t,t'\rightarrow 0$, but in general the limit can be taken such that
${t \over 2}~e^h \neq {t'\over 2}~e^{h'}$. For the sake of simplicity, we will take ${t \over 2}~e^h~=~{t'\over 2}~e^{h'}\rightarrow 1$. It will be obvious to the reader that the $G_0(z)$ for the more general anisotropic case can be obtained from the expressions below by a trivial rescaling. 

So, ``just for fun," we will evaluate
\beqra\label{one}
G_0(z) &= &\int ~{d \theta \over 2 \pi }~ {d \theta ' \over 2 \pi} ~{1
\over z-e^{i \theta}-e^{i \theta '}} \nonumber \\
&= &\int ~{d \theta \over 2 \pi }~{\theta \left(\mid z-e^{i \theta} \mid^2
-1 \right) \over z  ~-e^{i \theta}}
\eeqra
Similarly to our preceding discussion for the one dimensional case, for
$\mid z \mid ~>~2$, $G_0(z) ~=~ {1 \over z}$, and by (\ref{master2}) the density of eigenvalues vanishes there.
By rotational invariance, we can take $z$ 
to be real and equal to $x$ such that $0~<~x~<~2$, and obtain

\beq\label{goo}
G_0(x) ~=~ {1 \over 2 ~\pi~ x} ~ \int\limits_{{\rm arc ~cos~} {x \over 2}}^\pi ~ d 
\theta ~ {{\rm cos}~\theta ~-~x \over {\rm cos} ~ \theta ~-~ {1+x^2 \over 2x}}
\eeq
After some straightforward computation, we find for $\mid z \mid ~<~ 2$

\beq\label{goo2}
G_0(z) ~=~ {1 \over 2z} ~\left[1 ~-~{1 \over \pi} ~{\rm arc~cos}~{r \over 2} ~-~ {2 
\over \pi} ~{\rm arc ~tan} \left( \left({1-r \over 1+r}\right) ~ \sqrt{{2+r \over 
2-r}} \right) \right]
\eeq
where $r~=~ \mid z \mid$. The density of eigenvalues is given by (\ref{master2}) to be

\beqra\label{rho}
\rho_0(r) &= &{1 \over \pi} ~ {\partial~G_0 \over \partial z^\ast} \nonumber \\
&= &{1 \over 4 \pi} ~ {1 \over r}~ {du \over dr} \nonumber \\
&= &{1\over \pi^2}~{1 \over r \sqrt{4 - r^2}}
\eeqra
where $u(r)$ is the function defined by the square bracket in (\ref{goo2}).
Note that $\rho_0(r)$ diverges as $r~\rta~2$ from below. The density of eigenvalues
can of course also be obtained directly from its definition
\beq
\rho_0(x,y) ~=~ \lag~\int ~d\theta ~ \delta \left(x~-~ Re ~( e^{i \theta}+e^{i \theta '})
\right) ~\delta \left( y ~-~ Im ~ ( e^{i \theta}+e^{i \theta '}) \right) \rag
\eeq

When we turn on the Cauchy random site energies, the density of eigenvalues is
then given by (\ref{M}) and (\ref{Ngen}). Since we have $\rho_0$ explicitly, we
can work out various quantities.  For example, the fraction of eigenvalues on
the real axis is, according to (\ref{wing}), given by (for $\gamma\le 2$)

\beq\label{nr}
f_{real}(\gamma)~=~{4\over\pi^2}~ \int^{\gamma}_{0} ~ dv~ \int^{\sqrt{4-v^2}}_{0}~ du ~ {1\over\sqrt{(u^2+v^2)(4-u^2-v^2)}}
\eeq
We can obtain the $u$ integral in terms of an elliptic function and then determine $f_{real}(\gamma)$ numerically.  We show in figure (6) a plot of
$f_{real}(\gamma)$.
\begin{figure}[h]
\epsfxsize=4in
\begin{center}
\leavevmode
\epsfbox{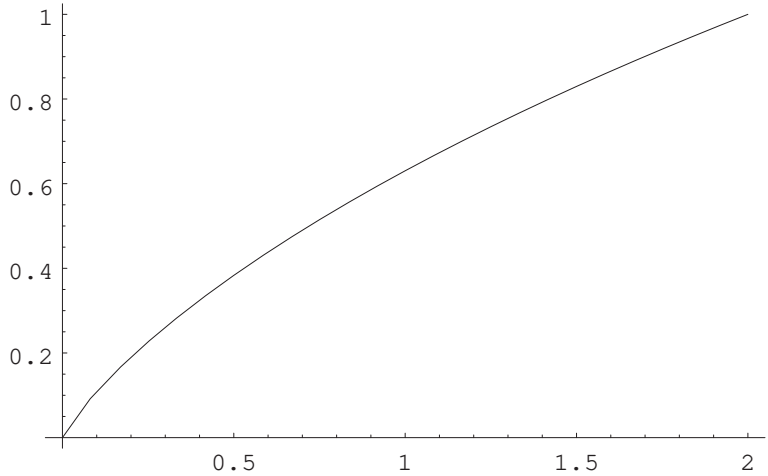}
\end{center}
\caption{The fraction of eigenvalues for $f_{real}$ as a function of randomness $\gam$.}
\label{fig6}
\end{figure}

We show some numerical results in figure(7) for two dimensional hopping with Cauchy
randomness.  Define the quantities $y_{outer} ~=~t'~{\rm sinh}~h' ~+~ t ~{\rm sinh}~h$ and
$y_{inner} ~=~t'~ {\rm sinh}~ h' ~-~ t~ {\rm sinh}~ h$.  For the parameters used in
figure(7), we have $y_{outer} ~=~ 0.789$ and $y_{inner} ~=~ 0.484$.  According to
(\ref{spectrum2}) and (\ref{spectrum}), $y_{outer}$ and $y_{inner}$ determine the
intersection of the outer and inner edges of the spectrum with the imaginary axis, for the range of the
parameter values chosen here. (The complete determination of the outer and
inner boundaries of the spectrum is straightforward but tedious, as it
involves the solution of a quartic equation. We will content ourselves here
with a remark about the simple case $t=t'$ in which $y_{inner}=t({\rm sinh}
h'-{\rm sinh} h)$. This expression is however not correct for all values of
$h$; in particular, not for $h=0$, as we can see geometrically. For $h$
small, we see from the geometrical construction ``ellipse $+$ ellipse" that
the correct expression is $y_{inner}=t({\rm sinh} ~h'~{\rm tanh} ~h')$. The
cross-over value of $h$ is then determined by ${\rm sinh}~ h={\rm sinh}~
h'~-{\rm sinh} ~h'~{\rm tanh}~ h'$.)

We have drawn in the values of $y_{outer}$ and $y_{inner}$ for the values
of $t, t', h,$ and $h'$ chosen here as the two horizontal lines  in figure
(7a); we see the outer and inner edges of the numerical spectrum indeed are equal to $y_{outer}$ and $y_{inner}$,
as it should be.  As discussed before (\ref{M}), when we turn on the Cauchy randomness,
the spectrum gets pushed towards the real axis by a distance $\gam$.  Thus, the critical
value of $\gam$ for the hole to disappear is given by 
\begin{figure}[htbp] 
\epsfxsize=4in
\begin{center} 
\leavevmode 
\epsfbox{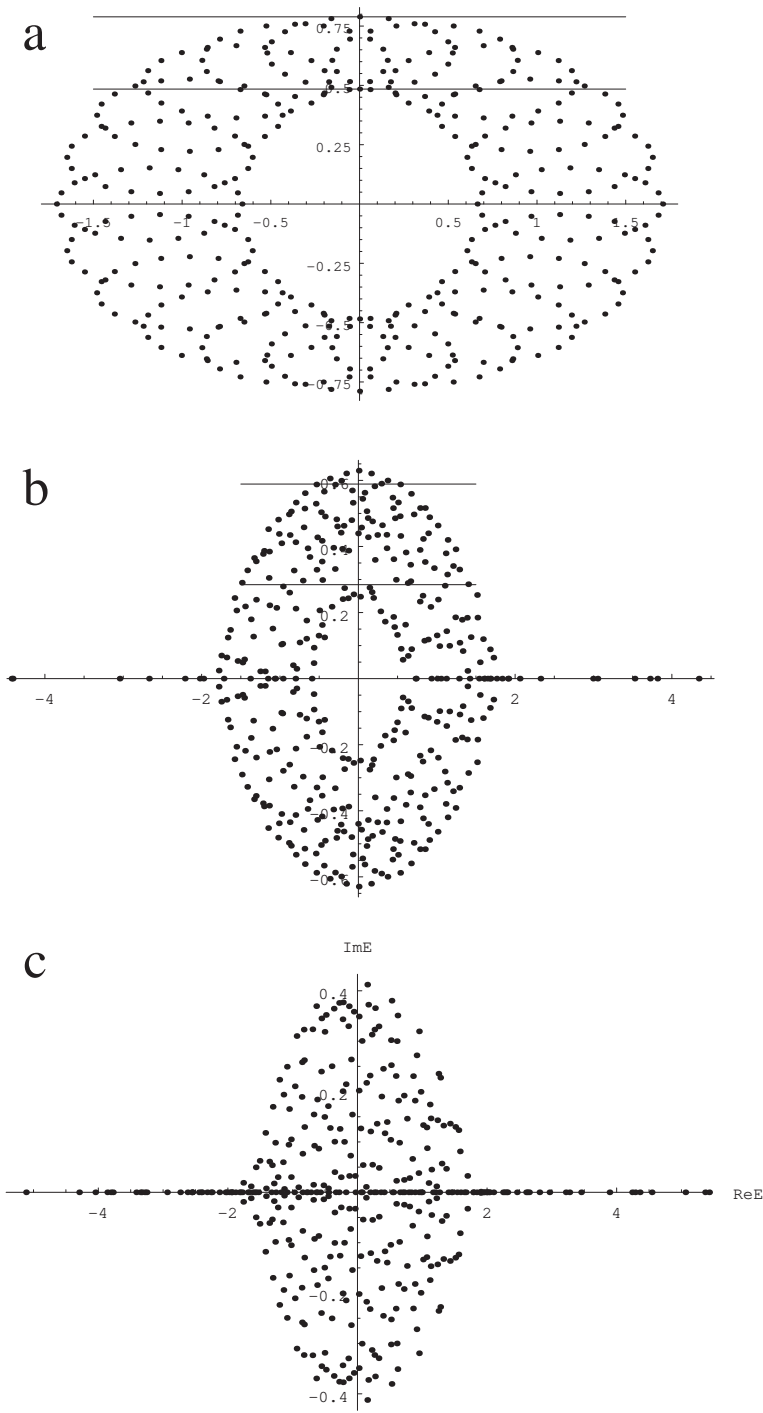} 
\end{center} 
\caption{The spectrum of $H$ on a two-dimensional $20 \times 20$ lattice with $t~=~0.5$, $h~=~0.3$, $t' ~=~ 1$, $h' ~=~
0.6$:  (a) without randomness (compare with figure 3a), (b) with Cauchy randomness of
strength $\gam ~=~ 0.2$, (c) with Cauchy randomness (b) with $\gam ~=~0.5$. Note the resemblance to a galactic cross section, as mentioned in the text.}  
\label{fig2}
\end{figure}

\beq\label{hole}
\gam_{c, ~hole} ~=~ t'~{\rm sinh}~h'~-~t~{\rm sinh}~h
\eeq
(for values of $t'$, $h'$, $t$, and $h$ such that a hole exists, and such that $y_{inner}$ is positive.) For figure (7), $\gam_{c,~hole} ~=~ 0.484$. We see from figure (7b) that at $\gam ~=~ 0.2$ the presence of the hole is clear. (The horizontal lines in figure (7b) represent ($y_{outer} ~-~ \gam$) and ($y_{inner} ~-~ \gam$).) In figure (7c), the hole has indeed disappeared. The fluctuations at finite $N$ can be studied using the density density correlation function defined in \cite{bz}.
\\

{\bf Acknowledgements}~~~
Part of this work was done in Paris and Tokyo, and I thank E. Br\'ezin and
S. Hikami respectively for their hospitality. I would also like to thank
C.K. Hu for his hospitality at the Statistical Physics Conference and N. Hatano for extensive discussions during the conference. I also benefitted from discussions with D. Nelson. I am grateful to M. Srednicki for helpful advice regarding numerical calculation. This
work was partly supported by the
National Science Foundation under Grant No. PHY89-04035.

\end{document}